\documentclass[twocolumn]{revtex4}
\usepackage{epsfig,subfigure,amssymb,amscd,amsmath,graphicx,amsfonts}
\usepackage{times}

\newcommand{\non}{\nonumber}

\newcommand{\bea}{\begin{eqnarray}}
\newcommand{\eea}{\end{eqnarray}}
\newcommand{\be}{\begin{equation}}
\newcommand{\ee}{\end{equation}}
\newcommand{\ba}{\begin{align}}
\newcommand{\ea}{\end{align}}

\newcommand{\ket}[1]{     |    \,    #1    \rangle}
\newcommand{\bra}[1]{  \langle #1  \,  |} 
\newcommand{\ZZ}{\mathbb{Z}}

\newcommand{\bk}{{\boldsymbol{k}}}

\newcommand{\bq}{{\boldsymbol{q}}}

\newcommand{\bs}[1]{ \boldsymbol{#1} }
\begin{document}

\title{Kaleidoscope of topological phases with multiple Majorana species}

\author{G. Kells$^{1,2}$, J. Kailasvuori$^{3,4}$, J. K.  Slingerland$^{2,5}$ and J. Vala$^{2,5}$}

\affiliation{$^{1}$  Institute for Theoretical Physics, Free University Berlin, Arnimallee 14, 14195 Berlin,  Germany \\$^{2}$  Department   of  Mathematical  Physics,  National University of Ireland, Maynooth, Ireland.  \\ $^{3}$ Max Planck Institute for the Physics of  Complex Systems, N\"othnitzer Str. 38, 01187, Dresden, Germany \\ $^{4}$ International Institute of Physics, Universidade Federal do Rio Grande do Norte, 59078-400 Natal-RN, Brazil\\$^{5}$ Dublin Institute for Advanced  Studies, School of Theoretical  Physics, 10 Burlington Rd, Dublin, Ireland. }

\begin{abstract}
\noindent
Exactly solvable lattice models for spins and non-interacting fermions provide fascinating examples of topological phases, some of them exhibiting the localized Majorana fermions that feature in proposals for topological quantum computing.  The Chern invariant $\nu$ is one important characterization of such phases. Here we look at the square-octagon variant of Kitaev's honeycomb model. It maps to spinful paired fermions and enjoys a rich phase diagram featuring distinct abelian and nonabelian phases with $\nu= 0,\pm1,\pm2,\pm3$ and $ \pm4$. The $\nu=\pm1 $ and $\nu=\pm3$ phases all support localized Majorana modes and are examples of Ising and $SU(2)_2$ anyon theories respectively. 
\end{abstract}


\date{\today}
\maketitle

\noindent
The study of topological quantum states has, in recent years, been boosted by proposals ~\cite{Kitaev03,Fre03,Nay08} to use nonabelian anyons  for topologically protected quantum computing.  Nonabelian anyons in two-dimensional systems were first proposed in the context of strongly  correlated systems, with  the Moore-Read Fractional Quantum Hall state~\cite{Moore-Read-1991} being the most famous example.  Recent years have seen the advent of nonabelian anyons in  effectively  non-interacting  systems. Proposals now exist for their realization in spinless chiral $p$-wave superconductors~\cite{Read-Green-2000} and spin lattices~\cite{Kitaev06}, at interfaces between an s-wave superconductors and a topological insulator~\cite{Fu-Kane-PRL-2008} or half metal~\cite{Duckheim2011}, and on 1-d semiconductor-superconductor heterostructures~\cite{Kitaev01,Oreg10,Lutchyn10}. In all of the above examples the nonabelian quasi-particles are realized by isolated, topologically protected Majorana fermions.

A necessary condition for isolated Majorana fermions is particle-hole symmetry of the type found in Bogoliubov-de Gennes (BdG) equations. A general understanding of where to find these conditions has been aided by the general symmetry classification of non-interacting gapped bulk systems \cite{Schnyder-etal-2008, Zirnbauer96, AltlandZirnbauer97,Kitaev-2009}. In particular, the symmetry class of 2d chiral $p-$ and $f-$wave superconductors (class D in \cite{Schnyder-etal-2008}) is topologically non-trivial and can host topological bulk and edge-modes. The topological invariant characterizing the ground state in 2d is the Chern number $\nu$, which maps to the set of all integers $\ZZ$.  However,in the chiral $p$-wave case, for example, only Chern numbers $\nu=0$ (abelian phase) and $\nu=\pm1$ (nonabelian phase)  are uncovered from the full set $\ZZ$.  This is to be contrasted with the symmetry class of the Integer quantum Hall effect  (class A), that lacks all symmetries. Here the $\ZZ$ invariant is fully realized (in theory) by filling some arbitrary number  $n$ of Landau level bands, which will be related to a Chern number  $\nu=\pm n$~\cite{TKNN82}. 

One of our motivations for searching for higher Chern numbers in systems with BdG symmetry is that phases with different odd $\nu$ host Majorana fermions of different topological properties. The Chern number dependence of the nonabelian braiding properties is discussed in  Kitaev's  seminal paper~\cite{Kitaev06}. A summary of the  \emph{$16$-fold way} of bulk anyon models ---so called because the braiding properties are determined by the value of  $\nu$ modulo $16$---is given in tables 1, 2 and 3 in that paper. The cases with odd $\nu$ (but not those with even $\nu$) all support localized Majorana modes at the vortices and thus have nonabelian exchange statistics \cite{Ivanov01}.  We will see that the lattice models studied in this paper realizes $9$ of these $16$ bulk orders, including $4$ of the cases with nonabelian anyons, in two parity-doublets $\nu=\pm 1$ and $\nu=\pm 3$. The only other example of Chern number $\nu=\pm 3$ that we are aware of occurs in the work of Mao and co-workers \cite{Mao2011}, which concerns an $f$-wave paired system of electrons moving in continuous spacetime. The Chern numbers $\nu=\pm 1$ and $\nu=\pm 3$ correspond to anyon models of the Ising and $SU(2)_2$ type respectively. However, the edge behavior of the system is not fully determined by this classification and will depend on the actual value of $\nu$ (not its residue modulo $16$). Nevertheless, by counting gapless edge modes for the various phases, we argue that the $\nu=\pm 1$ edges are described by chiral Ising CFTs (Conformal Field Theories), while the $\nu=\pm 3$ edges support chiral $SU(2)_2$ CFTs.

The model we study is  the square-octagon (4-8) lattice model discussed by Yang {\it et al.}~\cite{Yang07}. Like the triangle-dodecagon (3-12, ``Yao-Kivelson") lattice model ~\cite{YaoKivelson07} it is one of the possible generalizations in two dimensions  \cite{Yang07} of the original Kitaev honeycomb lattice model \cite{Kitaev06,Pachos07,Bas07,Lahtinen08,Lee07,Yu08,Feng07,Chen07b,Kells09b,Lahtinen11}. These exactly solvable 2d Ising-like spin models on trivalent lattices have turned out to be an ideal testing ground for topological properties predicted by the general symmetry arguments.  The honeycomb model maps to spinless chiral $p$-wave system, and exhibits an abelian ($\nu=0$) as well as nonabelian phases ($\nu=\pm 1$).   Yang and coworkers observed~\cite{Yang07} that the square-octagon variation has two abelian $\nu=0$ phases as well as the $\nu=\pm 1$ phases which open up near the critical boundary between the abelian ones. In none of the mentioned models were ground state (vortex-free) sectors with higher Chern numbers encountered. However, since the square-octagon model has a larger unit cell than the
model on the honeycomb lattice and can therefore be mapped to a fermionic paired
system with an internal pseudo-spin degree of freedom, we suspected that  it
may support the rich variety of phases found in spin-triplet
paired systems, see for example Ref.~\onlinecite{Volovik-book}.

\begin{figure*}[htb]
      \includegraphics[width=.80 \textwidth,height=0.40\textwidth]{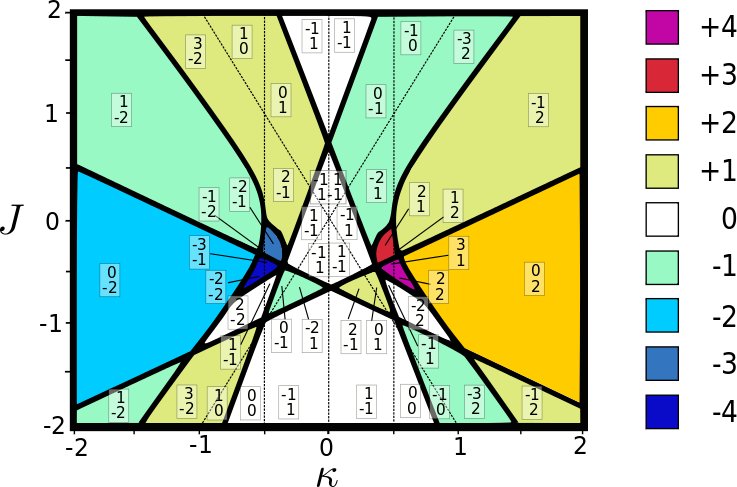}
      \caption{Phase diagram of the square-octagon model. In this figure $J_z=1$, $J=J_x=J_y$ and $\kappa=\kappa_1=\kappa_2$. The dark lines indicate phase transitions where the higher negative energy band and the lower positive energy band touch. Inside each phase the Chern numbers of the two negative bands are also indicated. The top number indicates the higher negative energy band, the one lying closest to $E=0$.   The dotted lines indicate where these negative energy bands touch. In this case the internal transfer of Chern integers does not change the total Chern invariant. All phases with total Chern number $\pm3,\pm4$ have topological orders for which we do not know of previous lattice realizations.}
      \label{fig:phase2}
\end{figure*}

We will see below that the the model supports an abundance of abelian and non-abelian phases with Chern numbers $0,\pm1,\pm2,\pm3,\pm4$ see Figure~\ref{fig:phase2}.  Using the fermionization method introduced in Ref.~\onlinecite{Kells09b}, we have also checked directly that this model can be mapped to a fermionic paired system with pairings of type s-, p-, d-, f- g- etc. along with spin-orbit couplings of corresponding winding numbers. In the absence of spin-orbit coupling---for example in spinless pairs---the correspondence between the Chern number and the angular momentum of the pairing is rather simple. A spinless chiral pairing of p-wave type corresponds in the weak-pairing phase to  $\nu=\pm 1$, a chiral f-wave pairing instead to $\nu=\pm3$ etc. (In the strong-pairing phase one would in both cases find $\mu=0$.)
The fact that we find phases with Chern numbers $\nu= \pm3$ among others concurs generally with the  f-wave pairing observed in Ref.~\onlinecite{Mao2011}. However, we stress that in the model considered by us---which as mentioned  contains  intricate couplings of pseudospin and momentum   with a  variety of  winding numbers---we have not been able to discern any simple relation between dominant pairing and the Chern number.


The outline of the paper is as follows. A general introduction to the spin model will be accompanied by a description of the fermionic solutions, focusing in particular on the vortex free sector, which contains the ground state.  We then describe the phase diagram and discuss how the edge spectrum, and indeed the bulk spectrum at phase boundaries reflect the Chern number difference between topological phases. We also provide details of how the contributions of each band in the model combine to give the total Chern number of each phase and describe how these contributions change as the bands cross or touch.  A general description of the edge spectra that occur between domains of the various topological phases (including vacuum) is then provided and we give an example where two domains with $\nu=0$ nevertheless have counter-propagating modes at the interface.  We also show that these two abelian $\nu=0$ phases can be described with toric code like hamiltonians, supported on different sub-lattice structures. 

\begin{figure}[tbh]
      \includegraphics[width=.40 \textwidth,height=0.35\textwidth]{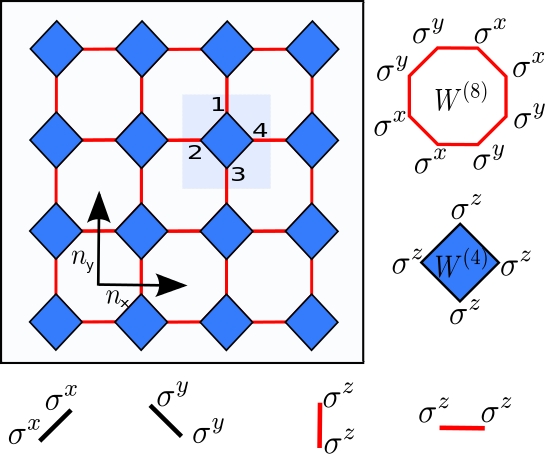}
      \caption{The square-octagon Kitaev lattice model. The horizontal links represent operators of the type $\sigma^x \sigma^x$, the vertical links represent operators $\sigma^y \sigma^y$ and the diagonal links represent the operators $\sigma^z \sigma^z$. Shown also are the fundamental unit cell with the sites inside it labeled by $n\in\{1,\ldots,4\}$, the lattice vectors $\bs{n}_x $ and $\bs{n}_y$, and the plaquette operators $\bs{W}^{\alpha}$.}
      \label{fig:SOlattice}
\end{figure}

The basic hamiltonian of the Kitaev-type trivalent spin model on the square-octagon (4-8) lattice is defined as  
\be
\label{eq:H}
H = - J_z \sum_{\text{z-links}}  \sigma^z \sigma^z  - J_x
\sum_{ \text{x-links} } \sigma^x \sigma^x  - J_y \sum_{\text{y-links}} \sigma^y \sigma^y 
\ee
where it should be understood that the $z$-links connect separate squares of the lattice and the $x$ and $y$-links within the squares are the horizontal and vertical slopes respectively, see Figure~\ref{fig:SOlattice}. We define a basic unit cell of the lattice around the four sites that make up a single square and label the sites inside such a cell by $n\in{1,2,3,4}$.  Each spin site on the lattice can thus be specified using the position vector $\bq$ of the unit cell and the index $n$. A unit cell and fundamental lattice vectors are shown in Figure~\ref{fig:SOlattice}. 

Within the model, as in honeycomb and 3-12 lattice models, there are closed loop symmetries that we can generate with overlapping link operators. The simplest of these are the octagonal $\bs{W}^{(8)}$ and square $\bs{W}^{(4)}$ plaquette operators. These are defined pictorially in Figure~\ref{fig:SOlattice}. The fact that the hamiltonian commutes with all plaquette operators implies that we may choose energy eigenvectors $\ket{n}$ such that $W^{(\alpha)}_{\bq}=\bra{n} \bs{W}^{(\alpha)}_\bq \ket{n}=\pm 1$. For example, if $W^{(4)}_{\bq} = -1$ then we say that the state $\ket{n}$ carries a square vortex at $\bq$. By the  vortex-sector we mean the subspace of the system Hilbert space with a particular configuration of vortices.  The vortex-free sector for example is the subspace spanned by all eigenvectors such that $W_{\bq}^{(4)} = 1$ and $W_{\bq}^{(8)} = 1$ for all $\bq$. The vortex-lattice sector, with $W_{\bq}^{(4)} = -1$ and $W_{\bq}^{(8)} = -1$ for all $\bq$, was shown in~\cite{Bas09} to have a Fermi-surface and is thus a spin metal, like certain sectors of the 3-12 lattice variation~\cite{Tikonov10}.
 
\begin{figure}[htb]
      \includegraphics[width=.40 \textwidth,height=0.25\textwidth]{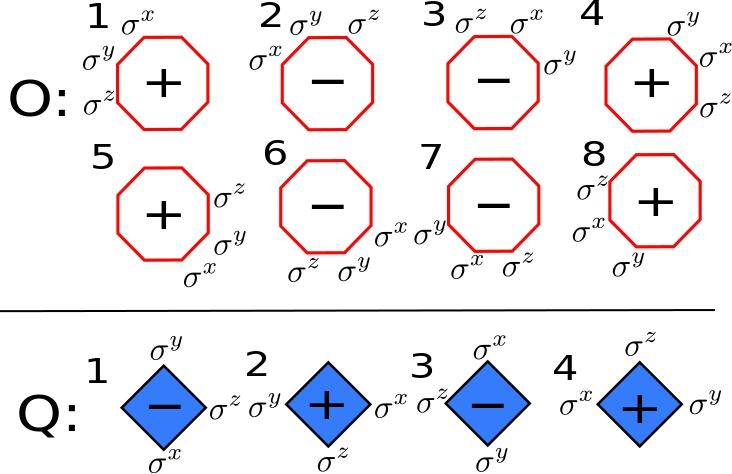}
      \caption{Time-reversal symmetry breaking terms. There are 8 octagonal (O) terms and 4 square (Q) terms. We construct each term using the overlapping products of two adjacent $\sigma^{\alpha} \sigma^{\alpha}$  terms. To fix the sign of each term, we use a clockwise ordering for the adjacent link operators. The resulting signs are noted inside the plaquettes. The O terms are added to the hamiltonian with an overall coupling constant $2\kappa_1$ and the Q terms with an overall coupling $\kappa_2$}
      \label{fig:Tbterms}
\end{figure}

In this type of system, it is possible to break time reversal symmetry explicitly, and still preserve the solvable nature of the system, by adding into the hamiltonian three site interactions of the form  $\sigma_i^\alpha \sigma_j^\beta \sigma_k^\gamma$. To find these terms we take overlapping products of the adjacent links of the original hamiltonian.  These terms are shown in Figure~\ref{fig:Tbterms}. 
Crucial to obtaining the rich phase structure shown in Figure~\ref{fig:phase2} is the sign we place in front of each term. Following~\cite{Yang07} we use the convention that the individual links are arranged in a clockwise order around each plaquette, which yields the signs shown in the figure. Two overall coupling constants $\kappa_1$ for the terms on square plaquettes and $\kappa_2$ for the terms on octagonal plaquettes allow us to vary the strength and nature of these $T$-breaking terms.

To solve the model exactly one converts each vortex sector of the model to a quadratic fermionic system. There are a number of different techniques that do this, each with their own distinct advantages. For the purposes of calculating the fermionic spectra it is easiest to use the Majorana representation and conventions for this model used in~\cite{Yang07,Bas09}. It is important to note however that this methodology means we artificially enlarge the Hilbert-space and thus should project back to the physical subspace in order to calculate wavefunctions~\cite{Kitaev06}.  Indeed, for the purpose of calculating wavefunctions it is arguably easier to use a Jordan-Wigner type fermionization procedure~\cite{Feng07,Chen07b}.  We have elsewhere given details on a representation that uses a special Jordan-Wigner type string to resolve the fermionic vacuum as Toric code stabilized states~\cite{Kells09b}. Ground states calculated with this methodology are BCS ground state wave functions over these Toric Code vacua and the full set of excited states can be described in terms of BdG quasi-particles over these ground states. 

We now focus on the vortex free sector, using the Majorana fermionic representation of ref.~\onlinecite{Yang07}. After this fermionization, the Hamiltonian can be written as follows
\be
H= \frac{1}{2} \sum_\bk A^\dagger_\bk M(\bk) A_\bk 
\ee
where $A^\dagger_\bk = (a_{\bk,1}, a_{\bk,2}, a_{\bk,3}, a_{\bk,4})$ and the operators $a_{\bk,n}$ are the Fourier transformed Majorana fermion creation/annihilation operators $a_{\bk,n}= \frac{1}{\sqrt{N}} \sum a_{\bq,n} e^{i \bk \cdot \bq}$. The indices $n$ label the four sites within a unit cell (see Figure~\ref{fig:SOlattice}), $\bq $ is the center-of-mass coordinate of the unit cell, and $a_{\bq,n}$ is the operator which creates or annihilates a Majorana fermion at the site labeled by $(\bq,n)$.
The $4\times4$ matrix $M$ can be decomposed as $M=M_J+M_{\kappa}$ where
\begin{equation}
M_J = 2 i\left[ \begin{array}{cccc}0&-J_x&J_z e^{-i k_y}&-J_y\\
 J_x&0&-J_y&-J_z e^{i k_x}\\
 -J_z e^{i k_y}&J_y&0&-J_x\\
 J_y&J_z e^{-i k_x}&J_x&0  \end{array} \right] 
\end{equation}
and
{\small
\begin{equation}
\frac{M_\kappa}{2 i \kappa_2}=\left[ \begin{array}{cccc}  
0&e^{\!-\!i k_x}\! -\!e^{\! -\!i k_y}&\! -\!2 \frac{\kappa_1}{\kappa_2}&e^{i k_x}\! -\!e^{\! -\!i k_y}\\
       e^{i k_y}\! -\!e^{i k_x}&0&e^{i k_x}\! -\!e^{\! -\!i k_y}&\! -\!2\frac{\kappa_1}{\kappa_2}\\
2\frac{\kappa_1}{\kappa_2} & e^{i k_y}\! -\!e^{\! -\!i k_x}&0 &e^{i k_x}\! -\!e^{i k_y}\\
e^{i k_y}\! -\!e^{\! -\!ik_x}&2 \frac{\kappa_1}{\kappa_2}&e^{\! -\!i ky}\! -\!e^{\! -\!i k_x}&0 \end{array} \right].
\end{equation}
}
The Chern invariant can be calculated as 
\be 
\nu = \frac{1}{2 \pi i} \int d^2 \bk Tr \left( P(k_x,k_y) \left[ \frac{\partial P}{\partial k_x} \frac{\partial P}{\partial k_y}- \frac{\partial P}{\partial k_y} \frac{\partial P}{\partial k_x} \right] \right)
\label{eq:Chern}
\ee
where $P$ is the projection onto the negative energy eigenvectors of $M$. For multi-banded systems one can calculate a Chern integer for each band by using the same formula, but with $P$ replaced by the projection to just that band. Each negative energy band then contributes its own integer to the total Chern invariant.  The Chern integers of the completely positive energy bands are just $-1$ times the analogues for the negative energy bands and therefore contain the same information.

\begin{figure}[htb]
      \includegraphics[width=.45 \textwidth,height=0.55\textwidth]{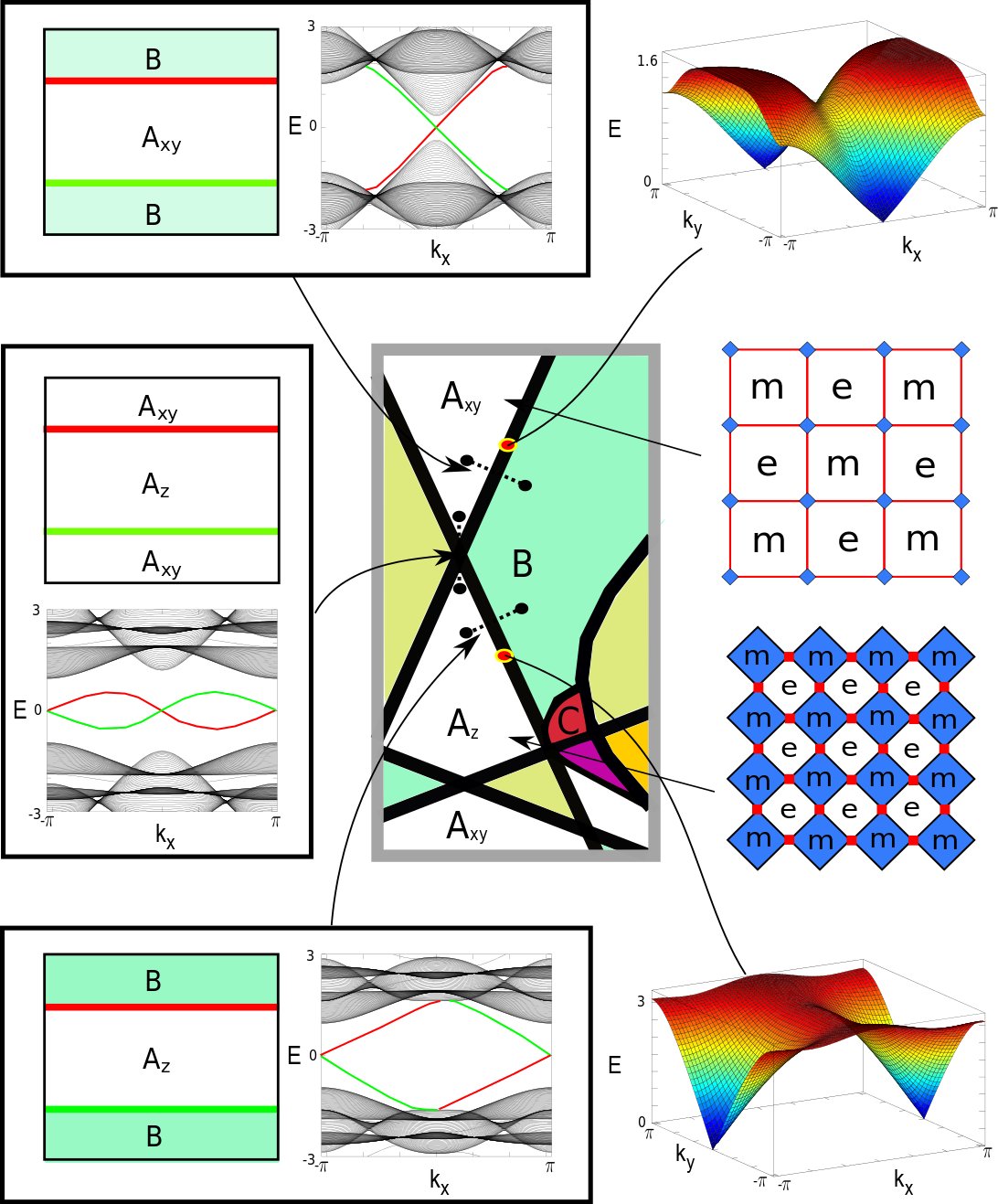}
      \caption{The central figure shows an excerpt of the central region of the phase diagram in Fig.~\ref{fig:phase2}. The domain wall edge spectra between two phases reflect all the bulk phase transitions that occur in the phase diagram along a continuous line connecting the two phases. Low energy edge modes associated with the red edge shown in red (in green for the green edge). 
      We show here the bulk phase diagram at a $A_{xy}-B$ and a $A_z -B$ transition and how it is reflected in the domain wall edge spectra. One can also see here how the representative edge spectra from $A_{xy}-B$ edge and $A_{z}-B$ edges are in some sense combined at an $A_{xy}-A_z$ edge.  The $A_{xy}$ and $A_z$ phases can both be perturbatively mapped to Toric Code type systems that are supported on different sublattices. }
      \label{fig:E6}
\end{figure}

A change in the  Chern number  (\ref{eq:Chern}) indicates a phase transition (the reverse is not always true as Figure~\ref{fig:E6} illustrates). Changes in Chern number can occur when the positive and negative energy bands touch at one or several points. We can thus explore the phase diagram of the model by mapping out the Chern numbers and the gap as a function of the parameters $(J_x,J_y,J_z,\kappa_1,\kappa_2)$. Clearly we can scale one of these parameters to $1$ (as long as it was non-zero to start with) and this means we have a 4-dimensional parameter space to deal with. We will present the full structure of the 4D phase diagram in a
separate paper~\cite{Us_in_prep}. Here, we focus on a two-dimensional slice through parameter space, setting $J_z=1$, $J=J_x=J_y$ and $\kappa_1=\kappa_2$.  In Fig.~\ref{fig:phase2} we show the phases and Chern numbers in this slice. 

The information in Figure~\ref{fig:phase2} suggests that we can think of the topological phase transitions as an exchange of the Chern integers between positive and negative energy eigenmodes. Band touching within the negative energy bands also comes with a transfer of Chern integers, but here the total Chern invariant is unchanged. In the model in question the two negative energy bands touch whenever $\kappa=0$, $\kappa=\pm J_z/2$ or $\kappa=\pm J/2$.  These internal `transitions' are also indicated in Figure~\ref{fig:phase2}.

The dispersion relation for the energy $E_{k_x,k_y}$ of the model can be obtained exactly, but is complicated and we will not write it down here. However, the critical lines shown in Figure~\ref{fig:phase2} have simple analytical descriptions. We find that in all these cases the gap closes linearly and that the number of isolated $E_{k_x,k_y}=0$ points at a given transition dictates how the Chern invariant jumps between phases. For example, the critical lines of positive slope in the diagram, $J=\frac{1}{2}(4 \kappa \pm \sqrt{2}(J_z+2 \kappa))$ correspond to single zeros of the energy at $(k_x,k_y)=(0,\pi)$. The same is true for the two lines of negative slope, $J=\frac{1}{2}(-4 \kappa \pm \sqrt{2}(J_z-2 \kappa))$ which have zero energy at the point $(k_x,k_y)=(\pi,0)$. 

The two hyperbolas in the diagram are given by $\kappa=\pm \sqrt{J_z^2+2 J^2}/2$. In both cases, the bands touch when $(k_x,k_y)=(0,0)$ and when $(k_x,k_y)=(\pi,\pi)$ so that $\nu$ jumps by $2$ over these curves. The upper critical curve surrounding the $\nu=\pm3$ phases and the lower critical curve surrounding the $\nu=\pm4$ phases have the positive and negative energy bands touch simultaneously at $4$ points in momentum space. These curves are given explicitly by $\kappa= \pm \frac{1}{4}(1 + 2J \pm \sqrt(1+4J+12 J^2))$. The curve separating the $\nu=-4$ phase from the $\nu=0$ phase and the curve separating the $\nu=-1$ phase from the $\nu=+3$ phase have each four zero-modes of the form  $(\pi,k_{y1}),(\pi,k_{y2}),(k_{x1},0)$ and $(k_{x2},0)$ with $k_{x1},k_{x2},k_{y1},k_{y2}\notin \{0, \pi\}$. Similarly, the curves separating the $\nu=4$ phase from the $\nu=0$ phase and the $\nu=+1$ phase from the $\nu=-3$ phase have $E_{k_x,k_y}=0$ at $(0,k_{y1}),(0,k_{y2}),(k_{x1},\pi)$ and $(k_{x2},\pi)$.

As proved in~\cite{Kitaev06}, each vortex in the odd Chern number phases is accompanied by a single localized zero energy Majorana fermion excitation (assuming the vortices are well-separated). These localized Majorana fermions in the different phases with odd $\nu$ make the vortices into nonabelian anyons and we obtain here $4$ out of the $8$ different types nonabelian anyons which occur in Kitaev's $16$-fold way for free fermionic bulk topological orders (in 2 pairs of types related by parity). Specifically, the $\nu=\pm 1$ modes correspond to the so called  Ising anyon theory and the $\nu=\pm3$ modes correspond to the  $SU(2)_2$ anyon theory. These zero energy excitations give rise to a ground state degeneracy and allow for the possibility of nonabelian statistics and fault tolerant quantum computation~\cite{Kitaev03,Fre03,Nay08}. As far as we know, this is the first time that different types of nonabelian anyons carrying Majorana modes have been found in a lattice model. We find zero energy modes at vortex excitations also in some of the abelian phases of the model, whose topological orders cover $5$ of the $8$ abelian possibilities. In these cases however, the modes are full Dirac modes (consisting of two Majorana modes) and are thus not topologically protected from local error processes. We will discuss one of these Dirac modes briefly below.

The multitude  of topological phases in the model allow one to set up a variety of interesting domain wall scenarios. The general result of Hatsugai~\cite{Hatsugai93} always applies here: The net number of left movers minus the number of right movers is equal to the difference in Chern number of the phases on either side of the boundary. In the simplest cases, for example for open boundaries, where the topological medium just ends, we find that this is usually satisfied in the simplest possible way; the edge theory is chiral and has a number of zero modes equal to $|\nu|$. This supports the idea that the open edge theories of the $\nu=\pm 1$ and $\nu=\pm 3$ phases are Ising and $SU(2)_2$ CFTs, the simplest CFTs which satisfy the bulk-boundary correspondence. 
 
\begin{figure}[htb]
     \includegraphics[width=.4 \textwidth,height=0.5\textwidth]{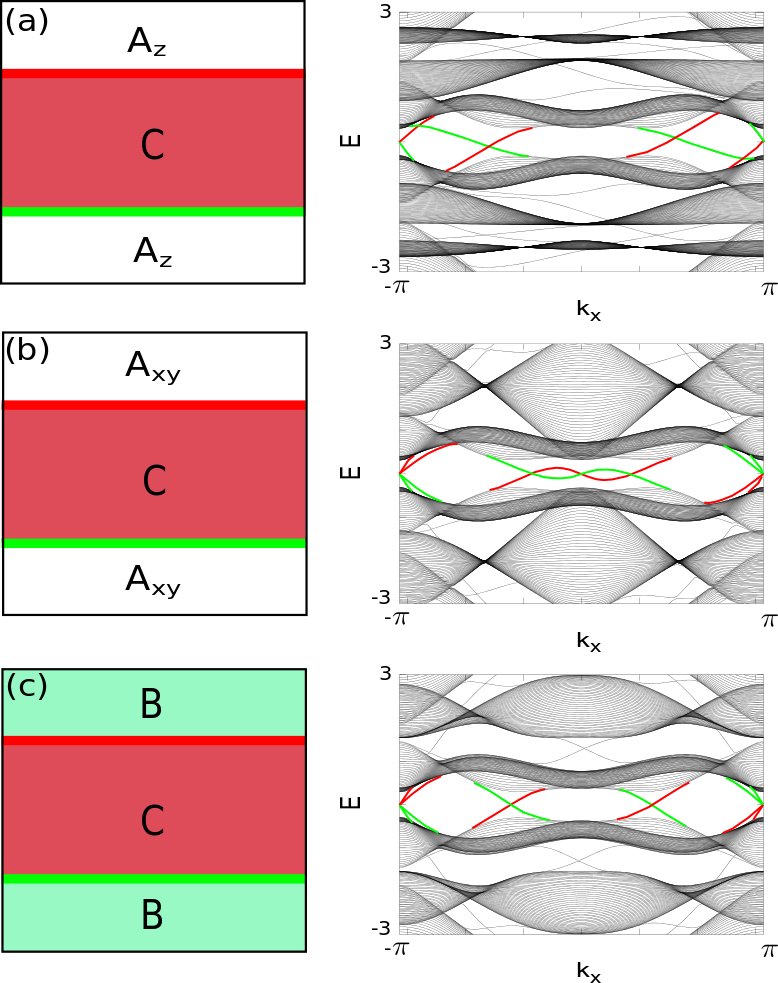}
      \caption{The edge spectra (a) between the C ($\nu=+3$) phase and the $A_z$ ($\nu=0$) phase, (b) between the C ($\nu=+3$) phase and the $A_{xy}$ ($\nu=0$) phase  and (c) between  the C ($\nu=+3$) phase and the $B$ ($\nu=-1$)  phase}
      \label{fig:EdgesC}
\end{figure}

For edge theories on domain walls the same is often true but not always. More generally we find that the number and type of zero modes that exist on a domain wall always reflects the critical bulk theory that exists on the phase boundary between the phases on either side of the wall.  If one examines the edge spectra in a  system with  one phase domain below another  in such a way that the translational invariance in the $x$ direction is maintained, one finds that the wall contains a band of states that intersect the $E=0$ line at value of $k_x$ that are consistent with the $k_x$ values of the zero modes that occur at the bulk phase transition. If the two phases are separated in the phase diagram by more than one phase boundary, then the edge will reflect all the phase boundaries crossed by an interpolating path. Examples of this behavior are given in Figures~\ref{fig:E6} and~\ref{fig:EdgesC}.

Careful inspection of the phase diagram reveals a number of situations where phases with the same Chern number touch at the intersections and touchings of critical lines. In these cases, we can have domain wall edge states between phases at the same Chern number. Such domain wall edge states are simply inherited from the boundaries between phases with different Chern numbers that meet at these multicritical points and hence they are stable against perturbations which do not remove the multicritical points. However, some of the multicritical points and corresponding edge states can be easily removed by local perturbations to the Hamiltonian, while others are topologically protected, that is, stable against any local perturbation which preserves particle-hole symmetry. We now give an example of each case. Following Yang and coworkers~\cite{Yang07} we call the phase around the $\kappa=0,|J|<J_z/\sqrt{2}$ domain the $A_z$ phase and around the $\kappa=0,|J|>J_z/\sqrt{2}$ line the $A_{xy}$ phase.  
It turns out that we can make the tetra-critical points that occur where the critical lines of positive slope cross and where the critical lines of negative slope cross vanish by varying $J_y$ away from $J_{x}$, in the process connecting the $\nu=0$ phases on the left and right to the $A_{xy}$ phase.   
In contrast, the tetra-critical points between the $A_{xy}$ and $A_z$ phases are robust features that cannot be so easily removed.  
In Figure~\ref{fig:E6} we show the edge spectrum for two horizontal domain walls between  $A_z$ phase and  the upper $A_{xy}$ phase. We see that on a single edge we have a single band of both left and right moving states that intersects $E=0$ at $k_x=0$ and $k_x=\pi$.
We now argue for the stability of these edge bands.

First of all, if the system is wide enough in the $y$-direction, the edge modes on opposite edges in Fig.~\ref{fig:E6} decouple exponentially and tunneling between the edges cannot lift the zero modes. Backscattering between the leftmoving and rightmoving zero modes on each edge could in principle lift these modes, but this would violate particle-hole symmetry. This is easiest to see for perturbations which preserve translational symmetry. In this case, because we have $E(k)=-E(-k)$ the modes at $k_x=0$ and $k_x=\pi$ must remain at zero energy. If translational symmetry is broken (for example by disorder), a slightly more delicate argument is needed. In this case $k_x$ is no longer a good quantum number, but if we imagine that the disorder was introduced adiabatically, we may still label the energy eigenstates by the former $k_x$-eigenvalues. The disorder may now make the energy of the $k_x=0$ and $k_x=\pi$ states nonzero, but in order to preserve particle hole symmetry, it must still preserve the property that every state at positive energy has a partner at negative energy. Therefore, in the continuum limit, where $k_x$ takes continuous values and energy is a continuous function of $k_x$, there will always be zero modes on the domain wall, though not necessarily at $k_x=0$ and $k_x=\pi$. In fact, there must always be a nonzero even number of zero modes, since zero modes can adiabatically appear and disappear in even numbers only. 

Of course $k_x$ is discrete for any finite system size and these arguments do not guarantee the existence of modes with exactly zero energy, just increasingly soft modes for growing system size. This is in fact how the system behaves when there is a finite and even number of unit cells in the $x$-direction. However, if there is an odd number of unit cells in the $x$-direction, the zero energy modes at $k_x=0$ and $k_x=\pi$ actually belong to different topological sectors and therefore backscattering between the $k_x=0$ and $k_x=\pi$ modes would violate conservation of a topologically conserved quantum number. In this situation we can have a robust Majorana zero mode forming on an interface between $\nu=0$ phases even at finite system size. 

It is gratifying that both Abelian phases in the model can be analysed using degenerate perturbation theory.  For the $\nu=0$ phase found in the center of the figure (the isolated z-link limit ($J_z \gg (J_x^2+J_y^2)^{1/2}$) it was shown in~\cite{Yang07} that the low-energy effective system, at the 4th order of perturbation theory, is that of an abelian toric code on a square lattice, 
\be 
\label{eq:HTSO}
\non H^{(4)}_z = - E_0 -  \frac{J_x^2 J_y^2}{16|J_z|^3} \sum_\bq \bar{\bs{W}}_\bq^{(4)} - \frac{ 5 J_x^2 J_y^2}{16 |J_z|^3} \sum_\bq \bar{\bs{W}}_\bq^{(8)}.
\ee
To the 4th order the constant terms can be calculated to be $E_0=N (|J_z|+ \frac{J_x J_y}{2|J_z|} + \frac{J_x^2 J_y^2}{32 |J_z|^3})$ where $N$ is number of z-links or fermions and is thus twice the number of unit cells.  In addition, we have found that such a mapping is possible also in the isolated square limit ($J_z \ll (J_x^2+J_y^2)^{1/2}$). Each isolated square is a miniature spin chain with alternating X and Y links as given by eq.~\ref{eq:H}.  In the absence of a vortex on the plaquette, it has a doubly degenerate ground state. This degenerate subspace acts as an effective spin in much the same way as the ground states of the isolated z-dimers above do. Perturbing away from the fully isolated regime, we can derive, at the 4th order, the  effective hamiltonian
\be
\label{eq:HTO}
H^{(4)}_s = - E_0 -  \frac{5 J_z^4}{8 r^3}\sum_\bq \bar{\bs{W}}_\bq^{(8)}, 
\ee
where $r = 2 (J_x^2+J_y^2)^{1/2}$ and $E_0=N (r+\frac{J_z^2}{r} +\frac{J_z^4}{4 r^3})$. In a suitable basis the operators $\bar{\bs{W}}_\bq^{(8)}$ are of the form $\bar{\sigma}^y_l \bar{\sigma}^y_r \bar{\sigma}^z_u \bar{\sigma}^z_d$. Here the subscript labels indicate left, right, up and down respectively and  the operators $\bar{\sigma}$ operate on the low energy anti-periodic sector of the isolated $4$-spin  chain, see Fig.~\ref{fig:E6}. In this regime, excitations on the octagonal plaquettes can be interpreted as electric $(e)$ and magnetic $(m)$ particles, in a checkerboard pattern as in ref.~\cite{Kitaev06}.  We see then that we have a toric code type phase, but on a different lattice than that found for the isolated z-link limit

Further distinctions between these Abelian phases can be made if we examine the system's square vortex excitations. Note first that the spectrum of the $4$-spin with a vortex excitation is four fold degenerate. This degeneracy is not lifted when we perturb away from the isolated square limit and so we can infer that in this $A_{xy}$ phase there exists a single massless Dirac fermion mode in the presence of a square-vortex excitation. It is important to note that this degeneracy is dependent on the fact that the chain is in an exact eigenstate of the square plaquette operator and hence it is not topologically protected (it can be broken by local perturbations).   

The domain wall between the abelian $A_z$ and $A_{xy}$ phases at $\nu=0$ is not the only one which shows counter-propagating zero modes. To illustrate this fact, we include figure~\ref{fig:EdgesC} which shows that domain walls involving the nonabelian $C$ phase, at $\nu=3$, exhibit similar phenomena.  The $C-A_z$ ($C-B$) boundary spectra shown behave as would be naively expected, with 3 (4) zero modes of the same chirality on each of the respective edges (shown in red and green). However, the $C-A_{xy}$ boundary actually retains the zero modes at $k_x=0$ and $k_x=\pi$ that comes from either taking the path $C-B-A_{xy}$ or $C-A_{z}-A_{xy}$ through the phase diagram.

In conclusion, in this paper  we have seen that the square-octagon  lattice variation of the Kitaev honeycomb model has a startling richness of topological phases. In particular it has two essentially different types of nonabelian anyons at Chern numbers $\nu=\pm1$ and $\nu=\pm 3$ and it realizes $9$ of the $16$ bulk topological orders allowed by Kitaev's $16$-fold way. We demonstrated how the bulk phase diagram of the system can be used to predict the behavior of the edge spectra between different topological domains. We encounter various situations where phases with the same Chern number have a non-zero (but even) number of low energy edge bands due to the interface. An examination of the perturbation theory of the abelian phases reveals that distinct abelian phases can both be mapped to Toric Code type systems that are supported on different sub-lattice structures.  

The perhaps most interesting aspect of the model is that it gives an example of a system with  Majorana modes that seizes the possibility of higher Chern numbers allowed from general symmetry classifications of Altland and Zirnbauer~\cite{Zirnbauer96,AltlandZirnbauer97}. Compared to the Kitaev honeycomb model that maps to spinless paired fermions, the pseudo fermion spin in the square-octagon  model doubles the number of bands and from this simple doubling of bands we could expect Chern numbers from 0 up  to $\pm2$. The real surprise is the occurrence of  $\nu=\pm 3,  \pm4$ which are beyond the expected doubling.  In particular,  we have found  instances where a single band can contribute with $\nu=\pm 3$. 

A deeper understanding of how in detail the multi-band situation allows for increased Chern numbers is left for future study, but it may be imagined that, given a sufficient number of sites in the unit cell of the lattice, it should be possible to realize all $16$ possible bulk topological orders for gapped Majorana systems within the same trivalent spin models.  In this respect, some progress has also been made using Bloch
Hamiltonians derived from the vortex sectors of the Kitaev honeycomb model and related continuum systems, see for example Ref.~\cite{Gils2009}. It is also interesting to speculate on the situation for  gapped phases in  3d variations ~\cite{3dmodels}  of Kitaev type models. The winding number is in this case given by a Wess-Zumino-Witten term\cite{Schnyder-etal-2008}. As the number of bands is typically higher, at least four, it seems probable that winding numbers   other than $\nu=0, \pm 1$ could be found.  The general symmetry arguments~\cite{Schnyder-etal-2008} provide a situation that is almost inverse to that in 2d.  Only one of the classes with a $\ZZ$ bulk comes with broken  T-symmetry, and this class---AIII---is probably less relevant to the Kitaev type models as it lacks particle-hole symmetry. The other two classes  with  $\ZZ$  bulks, namely DIII and CI, require particle-hole symmetry  {\it and} time-reversal symmetry. Nevertheless, it remains to be seen whether  higher winding  numbers can actually be realized, possibly by extending these T-invariant models with, for example, additional 4-spin interactions.  

\vspace{5mm}
G.~K.~was supported by the Alexander von Humboldt foundation. J.~K.~is grateful to R.~Moessner  for comments on the first draft and to R.~Roy for discussions. J.~K.~S.~was supported by Science Foundation Ireland Principal Investigator award 08/IN.1/I1961. J. V. was supported by Science Foundation Ireland through the President of Ireland Young Researcher Award 05/YI2/I680 and the Principal Investigator award 10/IN.1/I3013.

\end{document}